\documentclass[12pt]{iopart}
\usepackage[colorlinks=true,citecolor=blue,linkcolor=magenta]{hyperref}
\usepackage{graphicx}
\usepackage{bm}
\usepackage{dsfont}
  \expandafter\let\csname equation*\endcsname\relax
  \expandafter\let\csname endequation*\endcsname\relax
\usepackage{amsmath,amssymb}
\usepackage{color,psfrag}
\usepackage{pstool}
\usepackage{array}
\usepackage{longtable}
\usepackage{braket}
\usepackage{wrapfig}
\usepackage{bbm}
\usepackage{color}
\usepackage{mathrsfs}
\usepackage[export]{adjustbox}

\usepackage{multirow}
\usepackage{lipsum}

\begin{document}
\title{Optimization of photon storage fidelity in ordered atomic arrays}
\author{M.~T. Manzoni$^{1}$, M. Moreno-Cardoner$^{1}$, A. Asenjo-Garcia$^{1,2,3}$, J.~V. Porto$^4$, A.~V. Gorshkov$^{4,5}$, and D.~E. Chang$^{1,6}$}
\address{$^1$ ICFO-Institut de Ci\`encies Fot\`oniques, The Barcelona Institute of Science and Technology, 08860 Castelldefels (Barcelona), Spain.}
\address{$^2$ Norman Bridge Laboratory of Physics MC12-33, California Institute of Technology, Pasadena, CA 91125, USA}
\address{$^3$ Institute for Quantum Information and Matter, California Institute of Technology, Pasadena, CA 91125, USA}
\address{$^4$ Joint Quantum Institute, NIST/University of Maryland, College Park, Maryland 20742, USA}
\address{$^5$ Joint Center for Quantum Information and Computer Science, NIST/University of Maryland, College Park, Maryland 20742, USA}
\address{$^6$ ICREA -- Instituci\'o Catalana de Recerca i Estudis Avan\c{c}ats, 08015 Barcelona, Spain.}

\begin{abstract}
A major application for atomic ensembles consists of a quantum memory for light, in which an optical state can be reversibly converted to a collective atomic excitation on demand. There exists a well-known fundamental bound on the storage error, when the ensemble is describable by a continuous medium governed by the Maxwell-Bloch equations. However, these equations are semi-phenomenological, as they treat emission of the atoms into other directions other than the mode of interest as being independent. On the other hand, in systems such as dense, ordered atomic arrays, atoms interact with each other strongly and spatial interference of the emitted light might be exploited to suppress emission into unwanted directions, thereby enabling improved error bounds. Here, we develop a general formalism that fully accounts for spatial interference, and which finds the maximum storage efficiency for a single photon with known spatial input mode into a collection of atoms with discrete, known positions. As an example, we apply this technique to study a finite two-dimensional square array of atoms. We show that such a system enables a storage error that scales with atom number $N_\mathrm{a}$ like $\sim (\log N_\mathrm{a})^2/N_\mathrm{a}^2$, and that, remarkably, an array of just $4 \times 4$ atoms in principle allows for an error of less than 1\%, which is comparable to a disordered ensemble with optical depth of around 600.
\end{abstract}

Atomic ensembles constitute an important platform for quantum light-matter interfaces \cite{Hammerer2010a}, enabling applications from quantum memories \cite{Julsgaard2004,Choi2008,Liu2001,Lukin2001b} and few-photon nonlinear optics \cite{Murray2016,Pritchard2013a,Dudin2012,Peyronel2012a,Gorniaczyk2014,Tiarks2014} to metrology \cite{Kuzmich2000,Wasilewski2010,Leroux2010,Sewell2012}. In typical experiments, ensembles consist of disordered atomic clouds, with the propagation of light through them modeled phenomenologically by the Maxwell-Bloch equations \cite{Arecchi1965,McCall1967}. Within this description, the atoms are treated as a smooth density and the discreteness of atomic positions is ignored. In addition, spatial interference that can arise from light scattering is neglected, and the emission into directions other than the mode of interest is treated as an independent atomic process. Within this formalism, one can derive standard limits of fidelity for applications of interest -- for example, the storage error of a quantum memory scales inversely with the optical depth (D) of the ensemble \cite{Gorshkov2007}.

Recently, novel experimental platforms have emerged where it is possible to produce small ordered arrays of atoms \cite{Lester2015,Barredo2016,Endres2016,Haller2015,Greif2016}. Intuitively, one expects that strong interference in light emission can emerge, which renders inoperable the typical theoretical approaches to modeling light-atom interfaces. Theoretically there has been growing interest in novel quantum optical effects in arrays, such as subradiance \cite{Zoubi2010,Bettles2015,Bettles2016,Sutherland2016,Hebenstreit2017,Asenjo2017b,Facchinetti2016}, topological effects \cite{Perczel2017,Bettles2017}, and complete reflection of light \cite{Abajo2007,Bettles2016b,Shahmoon2017}. Indeed, it has already been shown numerically that an ordered one-dimensional array of atoms coupled to a nanofiber allows for a storage error exponentially smaller than the previously known bound \cite{Asenjo2017b}. In this work, the exponential scaling was observed by considering a fixed, spatial waveform for the optical pulse. However, two interesting questions that arise are (i) whether it is possible to develop a theoretical technique to bound the error, which takes fully into account the atomic positions and the interference of emission in all directions, and (ii) whether an improved scaling is possible for atoms in free space, as opposed to coupled to a photonic structure. These questions are affirmatively answered in our work.

In particular, we provide a construction that enables the maximum storage efficiency to be found, given the atomic positions and the desired spatial mode of light. This procedure is based upon solving the dynamics of a ``spin model", which encodes the multiple scattering and interference of light as it interacts with atoms, and then calculating the light emitted into the desired mode by an input-output equation. We show that the maximum efficiency is given by the maximum eigenvalue of a Hermitian matrix, whose elements are derived from the atomic positions and optical mode. While this technique is completely general, we apply it specifically to the case of a two-dimensional square array of atoms. In particular, it has recently been shown that an infinite array can in principle form a 100\% reflector for light \cite{Abajo2007,Bettles2016b,Shahmoon2017}, when the lattice constant $d$ is smaller than the resonant wavelength $\lambda_0$. While a mirror constitutes a ``passive" optical system, it is natural to ask whether this implies a 100\% success probability, if the system were functionalized into a quantum memory. For a finite array, we show that the minimum error decreases like $\epsilon\sim (\log N_\mathrm{a})^2/N_\mathrm{a}^2$ for storage from a Gaussian-like mode, and remarkably, that a $4 \times 4$ array in principle already enables an error below 1\%. 

\section{The spin model} 
The full dynamics of light emission and re-scattering of an arbitrary collection of atoms in free space, specified only by their discrete, fixed positions $\mathbf{r}_j$, can be related to an effective model containing only the atomic degrees of freedom and the incident field \cite{Gross1982,Asenjo2017,Dung2002,Buhmann2007,GW96,DKW02}. We first review this formalism for two-level atoms with ground state $\ket{g}$ and excited state $\ket{e}$, with the dipolar transition $\ket{g}-\ket{e}$ coupled with free space optical modes. Within the Born-Markov approximation, these modes can be integrated out to yield effective dynamics for the atomic density matrix $\hat{\rho}$, which evolves as $\dot{\hat{\rho}}=-(i/\hbar)[H,\hat{\rho}]+\mathcal{L}[\hat{\rho}]$, where the Hamiltonian and Lindblad operators read \cite{Gross1982,Asenjo2017,Dung2002,Buhmann2007,GW96,DKW02,Scully2008}
\begin{subequations}
\begin{equation}\label{ham}
H=H_\text{in}-\mu_0d^2_{eg}\omega_{eg}^2\sum_{j,l}\hat{\mathbf{d}}_j^*\cdot \text{Re}\{\mathbf{G}_0(\mathbf{r}_j,\mathbf{r}_l,\omega_{eg})\}\cdot \hat{\mathbf{d}}_l\,\sigma^{eg}_j\sigma^{ge}_l,
\end{equation}
\begin{equation}\label{lind}
\mathcal{L}[\hat{\rho}]=\frac{1}{2}\mu_0d^2_{eg}\omega_{eg}^2\sum_{j,l}\hat{\mathbf{d}}_j^*\cdot \text{Im}\{\mathbf{G}_0(\mathbf{r}_j,\mathbf{r}_l,\omega_{eg})\}\cdot \hat{\mathbf{d}}_l\,\left(2\sigma^{ge}_l\hat{\rho}\sigma^{eg}_j-\sigma^{eg}_j\sigma^{ge}_l\hat{\rho}-\hat{\rho}\sigma^{eg}_j\sigma^{ge}_l\right).
\end{equation}
\end{subequations}
Here $H_\mathrm{in}$ is associated with the input field that drives the atoms (which need not be specified for our purposes), $d_{eg}$ and $\hat{\mathbf{d}}_j$ are the dipole matrix element and unit atomic polarization vector associated with the transition, and $\sigma^{\beta\gamma}=\ket{\beta}\bra{\gamma}$ are atomic operators with $\{\beta,\gamma\} \in \{e,g\}$. $\mathbf{G}_0(\mathbf{r}_j,\mathbf{r}_l,\omega_{eg})$ is the electromagnetic Green's function tensor in free space, which is the fundamental solution of the wave equation and fulfills
\begin{equation}
\label{wave_eq}
\nabla\times\nabla\times{\mathbf G_0}(\mathbf{r},\mathbf{r}',\omega_{eg})-\frac{\omega_{eg}^2}{c^2}\,{\mathbf G_0}(\mathbf{r},\mathbf{r}',\omega_{eg})=\delta(\mathbf{r}-\mathbf{r}') \mathbf{I},
\end{equation}
where the curl is taken with respect to $\textbf{r}$. The Green's function explicitly takes the form  \cite{Novotny2006}
\begin{equation}
\label{Greens}
\mathbf{G}_0(\mathbf{r}_j,\mathbf{r}_l, \omega_{eg}) = \frac{e^{i k_0R}}{4\pi R}\bigg[\bigg(1+\frac{i k_0R-1}{k_0^2R^2}\bigg)\mathbf{I} + \frac{3-3i k_0R - k_0^2R^2}{k_0^2R^2}\frac{\mathbf{R}\otimes\mathbf{R}}{R^2}\bigg],
\end{equation}
where $R = |\mathbf{r}_j-\mathbf{r}_l|$ and $k_0=\omega_{eg}/c$ is the wavevector associated with the atomic transition frequency $\omega_{eg}$, with $c$ being the speed of light. We note that the local term [i.e., $\mathbf{G}_0(\mathbf{r}_j,\mathbf{r}_j, \omega_{eg})$] is divergent. This term is responsible for the Lamb shift and is incorporated into a renormalized resonance frequency $\omega_{eg}$. Physically, Eq.~\eqref{ham} describes the coherent exchange of atomic excitations mediated by photons. On the other hand, Eq.~\eqref{lind} describes the collective emission or dissipation of excited atoms, after integrating out the common reservoir of electromagnetic modes with which they interact (within the Born-Markov approximation).

Instead of solving the density matrix evolution as governed by the master equation, one can equivalently work within the stochastic wave function or ``quantum jump" formalism~\cite{Meystre2007}. In that case, the system is described by a wave function, which deterministically evolves under an effective, non-Hermitian Hamiltonian
\begin{equation}
\label{eff_Ham}
H_\mathrm{eff} = H_\mathrm{in} - \mu_0d^2_{eg}\omega_{eg}^2\sum_{j,l}\hat{\mathbf{d}}_j^*\cdot\mathbf{G}_0(\mathbf{r}_j,\mathbf{r}_l,\omega_{eg})\cdot \hat{\mathbf{d}}_l\,\sigma^{eg}_j\sigma^{ge}_l.
\end{equation}
This Hamiltonian captures both the coherent evolution of Eq.~\eqref{ham} and the last two terms of the Lindblad operator in Eq.~\eqref{lind}. In addition, one must also stochastically apply quantum jump operators to the wave function, to capture the population recycling terms $\sigma_{ge}^{l}\rho\sigma_{eg}^{j}$ of Eq.~\eqref{lind}. Formally, the jump operators of our system will consist of superpositions of $\sigma_{ge}^{l}$, i.e. atomic lowering operators, which physically encode the emission of a photon. In the following, we will be interested in initial states with just a single excitation; thus, any jump operator trivially takes the system to the ground state $\ket{g}^{\otimes N}$, where it cannot further evolve or contribute to observables of interest (e.g., the emission of a photon). Furthermore, the rate that jumps occur is exactly equal to the rate of population loss of the wave function evolving under $H_\mathrm{eff}$. Thus, in our case, jumps are effectively accounted for just by evolution under $H_\mathrm{eff}$ alone. Any loss of population from the single-excitation manifold implies that a corresponding population is building up in the manifold $\ket{g}^{\otimes N}\ket{1(r,t)}$, where all the atoms are in the ground state and a single photon is emitted in some spatial-temporal pattern. We next discuss how the photon-emission pattern and its overlap with a mode of interest can be calculated. 

\begin{figure}[t]
\includegraphics[width=13cm,right]{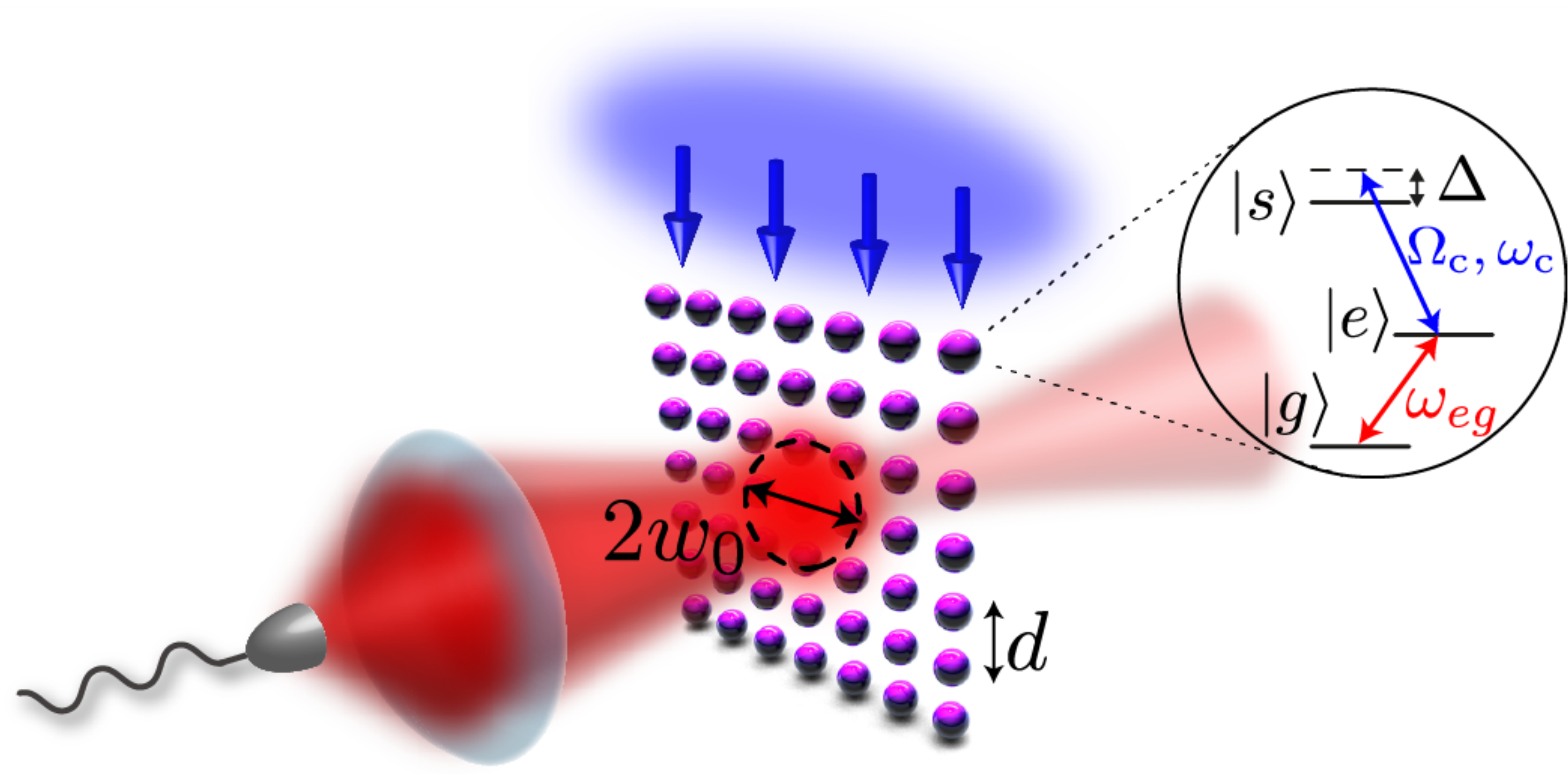}
\caption{Schematic of a quantum memory using a two-dimensional atomic array. An excitation initially stored in the $\ket{s}$-manifold is retrieved as a photon by turning on the classical control field $\Omega_c$ (blue arrows), which then creates a Raman scattered photon from the $\ket{g}-\ket{e}$ transition. The photon is detected in some given mode, illustrated here as a Gaussian beam.}
\label{fig:array}
\end{figure}

Given the evolution of the atomic state under $H_\mathrm{eff}$, any observables associated with the total field operator $\hat{\mathbf{E}}_\mathrm{out}(\mathbf{r})$ can be derived from the input-output relation \cite{Asenjo2017,Dung2002,Buhmann2007,GW96,DKW02}
\begin{equation}
\label{output}
\hat{\mathbf{E}}_\mathrm{out}(\mathbf{r}) = \hat{\mathbf{E}}_\mathrm{in}(\mathbf{r}) + \mu_0d_{eg}\omega_{eg}^2\sum_{j} \mathbf{G}_0(\mathbf{r},\mathbf{r}_j,\omega_{eg})\cdot \hat{\mathbf{d}}_j\sigma^{ge}_j.
\end{equation}
Formally, this equation states that the total field is a superposition of the incoming field and the fields emitted by the atoms, whose spatial pattern is contained in the Green's function. Equation \eqref{output} enables the field to be calculated at any point $\mathbf{r}$, based upon the evaluation of an atomic correlation function $\sim \mathbf{G}_0(\mathbf{r},\mathbf{r}_j,\omega_{eg})\cdot\hat{\mathbf{d}}_j\sigma_{j}^{ge}$ weighted by the Green's function. Evaluating the Green's function at each $\mathbf{r}$ and the corresponding atomic correlation function to construct the field everywhere can become tedious. However, in experiments one often cares about the projection of the field into a specific spatial mode, such as a Gaussian (see Fig.~\ref{fig:array}). It can be proven (see~\ref{ApA}) that this projection depends only on the amplitudes of the mode of the classical field $\mathbf{E}_\mathrm{det}(\mathbf{r})$ at the positions of the dipoles. We can thus define the quantum operator associated with the detector as
\begin{equation}
\label{overlap}
\hat{E}_\mathrm{det} = \hat{E}_\mathrm{det, in} + id_{eg}\sqrt{\frac{k_0}{2\hbar \epsilon_0F_\mathrm{det}}}\sum_j \mathbf{E}_\mathrm{det}^*(\mathbf{r}_j)\cdot \hat{\mathbf{d}}_j\sigma^{ge}_j,
\end{equation}
where $\hat{E}_\mathrm{det, in}$ is the input field in the detection mode and $F_\mathrm{det} = \int_{z=\mathrm{const}} d^2\mathbf{r}\,\mathbf{E}_\mathrm{det}^*(\mathbf{r})\cdot \mathbf{E}_\mathrm{det}(\mathbf{r})$ is a normalization factor. Here, the normalization is such that $\langle\hat{E}^\dagger_\mathrm{det}\hat{E}_\mathrm{det}\rangle$ represents the photon number per unit time emitted into the mode.

Before discussing the specifics of the retrieval efficiency, we would like to briefly discuss the validity of the Born-Markov approximation, which allows one to trace out the photonic degrees of freedom and arrive at an atomic master equation, as well as to write equations for the field operators that depend instantaneously on the atomic operators. This approximation is valid whenever (1) the photon bath correlations decay much faster than the atomic correlations and (2) retardation can be ignored. The first condition is obviously satisfied for atoms in free space, as the vacuum's Green's function has a frequency spectrum that is much broader than the atomic linewidth. Neglecting retardation in both the photon-mediated interactions between atoms and the field produced by the atoms requires the characteristic length $L$ of the atomic system to be much smaller than that of a spontaneously-emitted photon, which is $\sim c/\Gamma_0\leq 1$ m \cite{Scully2010,Ballestero2013,Shi2015,Guimond2016}, where $\Gamma_0=\mu_0\omega_{eg}^3d_{eg}^2/3\pi\hbar c$ is the single-atom spontaneous emission rate in vacuum. It should also be pointed out that for at most a single atomic excitation, the dynamics of atom-light interactions can readily be solved in an exact manner~\cite{Ballestero2013,Guimond2016,WSL04,Suttorp2004,YVR09}. In this regime of linear optics, the dynamics can be analyzed for each frequency component in the Fourier domain, exploiting the fact that different frequency components do not couple to one another. However, the spin model presented above has a natural extension to the multi-excitation case (e.g., studying the storage of multiple photons and their subsequent nonlinear interaction~\cite{Manzonidouglas,albrecht2017,Bienias2018}), whereas exact solutions are only available in a limited number of cases \cite{Pletyukhov2012, Shi2015, Mahmoodian2018}.

%%%%%%%%%%%%%%%%%%%%%%%%%%%%%%%%%%%%%%%%%%%%%%%%%%%%%%%%%%%%%%%%%%%%%%%%%%%%%%%%%%%%
%%%%%%%%%%%%%%%%%%%%%%%%%%%%%%%%%%%%%%%%%%%%%%%%%%%%%%%%%%%%%%%%%%%%%%%%%%%%%%%%%%%%

\section{The retrieval efficiency} 
The typical quantum memory scheme consists of an ensemble of three-level atoms where an additional metastable state $\ket{s}$ is coupled to the excited state $\ket{e}$ by a classical control field with Rabi frequency $\Omega_c(\mathbf{r},t)$ and detuning $\Delta$ from the transition frequency $\omega_{se}$ (see Fig.~\ref{fig:array}) \cite{Gorshkov2007}. While the state $\ket{s}$ is typically associated with another state in the ground-state hyperfine manifold, in our case this would deleteriously reduce interference effects in emission. For example, in storage where all atoms begin in $\ket{g}$, there is no interference pathway to suppress spontaneous emission into $\ket{s}$ once an incident photon excites an atom to $\ket{e}$. Thus, we assume that our atoms have no hyperfine structure and there is a unique ground state, as would be the case for bosonic Sr or Yb atoms, and that level $\ket{s}$ is a long-lived, higher-lying excited state. Dipole-dipole interactions on the $\ket{e}$-$\ket{s}$ transition have no effect, as they require at least two total excitations in the system. In the main text, we will furthermore take the conceptually simpler case where $\ket{e}$ is the unique excited state coupled to $\ket{g}$ (for concreteness, with polarization $\hat{\mathbf{d}}_j=\hat{x})$. A more realistic model with three excited states $\ket{e_{x,y,z}}$, providing an isotropic atomic response to light, is presented in \ref{ApC}, but the results qualitatively remain the same.

Instead of storage, it is mathematically more convenient to optimize the retrieval problem, in which an initial collective spin excitation $\ket{\psi(t=0)}=\sum_j s_j(t=0) \sigma_j^{sg} \ket{g}^{\otimes N_\mathrm{a}}$ is emitted as an outgoing photon on the $\ket{g}-\ket{e}$ transition via a Raman process facilitated by the control field $\Omega_c$. The initial state then evolves under the total Hamiltonian $H=H_\mathrm{eff}+H_\mathrm{c}$, where the Hamiltonian associated with the control field is $H_\mathrm{c}=\sum_j -\hbar\Delta\sigma^{ee}_j
+ \hbar\Omega_c^j(t)(\sigma_j^{es}+\text{H.c.})$ and $H_\mathrm{in}=0$ as there is no external field driving the $\ket{g}-\ket{e}$ transition in retrieval. We take a spatially uniform, but possibly time-dependent, control field [$\Omega_c^j(t)\equiv \Omega_c(t)$], although it is straightforward to generalize the following discussion to the case of a spatially varying control field. Then, for a given detection mode and atomic spatial configuration, we want to find the initial spin amplitude $s_j(0)$ that maximizes the retrieval efficiency. By time-reversal symmetry, the storage efficiency for an incoming photon in the same mode and for the same atomic configuration is identical, when optimized over the temporal shapes of the incoming photon and control field \cite{Gorshkov2007}. Writing the general state in time as $\ket{\psi(t)}=\sum_j (e_j(t)\sigma_j^{eg}+s_j(t)\sigma^{sg}_j)\ket{g}^{\otimes N_\mathrm{a}}$, the state amplitudes obey
\begin{eqnarray}
\label{eom}
\dot{e}_j&=&i\Delta e_j - i\Omega_c(t) s_j + i\Gamma_0\sum_{l}M_{jl} e_l, \\\label{eom_2}
\dot{s}_j&=& - i\Omega_c(t)e_j,
\end{eqnarray}
where the matrix $M_{jl} = 3\pi k_0^{-1}\hat{\mathbf{d}}_j^*\cdot\mathbf{G}_0(\mathbf{r}_j,\mathbf{r}_l,\omega_{eg})\cdot \hat{\mathbf{d}}_l$. While we explicitly consider the model above, we note that it is straightforward to add a number of other effects (e.g., decay of the $\ket{s}$ state or dephasing) into the analysis.

From Eq.~\eqref{overlap}, we can evaluate the expected total photon number $\eta=\int_0^{\infty} dt \langle\hat{E}^\dagger_\mathrm{det}(t)\hat{E}_\mathrm{det}(t)\rangle$ emitted into the detection mode. Assuming that the control field is turned on for long enough, it is guaranteed that one photon in total is emitted into all modes, and thus $\eta$ also represents the retrieval efficiency. Evaluating the atomic operators in Eq.~\eqref{overlap}, we find that
\begin{equation}
\label{efficiency}
\eta = \frac{S_{\lambda_0}\Gamma_0}{4F_\mathrm{det}}\,\sum_{j,l}\,E^*_jE_l\,\int_0^\infty dt\,e_j(t)e_l^*(t),
\end{equation}
where we have defined the local scalar field $E_j = \mathbf{E}_\mathrm{det}(\mathbf{r}_j)\cdot\hat{\mathbf{d}}^*_j$ at the atom positions, and $S_{\lambda_0} = (3/2\pi)\lambda_0^2$ is the resonant atomic optical cross-section  ($\lambda_0=2\pi/k_0$ being the resonant wavelength).

Equation \eqref{efficiency} can be simplified by noting that $M_{jl}$ in Eq.~\eqref{eom} is a symmetric complex matrix. Thus, if $M_{jl}$ is diagonalizable (as we numerically verify in our cases of interest), its eigenvalues $\lambda_\xi$ are complex and its eigenmodes $\mathbf{v}_\xi$ are non-orthogonal in the quantum mechanical sense, but obey the orthogonality and completeness conditions $\mathbf{v}^T_\xi\cdot\mathbf{v}_{\xi'} = \delta_{\xi\xi'}$ and $\sum_{\xi}\mathbf{v}_\xi\otimes\mathbf{v}^T_\xi = \mathbf{I}$ \cite{Asenjo2017}. Projecting the equations of motion into this basis results in $N_\mathrm{a}$ decoupled pairs of equations:
\begin{eqnarray}
\dot{e}_\xi &=&i(\Delta + \Gamma_0\lambda_\xi)e_\xi - i\Omega_c(t)s_\xi, \label{eom_retrieval_1}\\
\dot{s}_\xi&=& - i\Omega_c(t)e_\xi,
\label{eom_retrieval_2}
\end{eqnarray}
where $e_\xi = \sum_jv_{\xi,j}e_j$, $s_\xi = \sum_jv_{\xi,j}s_j$. Provided that the atomic excitation has left the system as $t\rightarrow\infty$, one can derive that
\begin{equation}
\int_0^\infty dt\,e_j(t)e^*_l(t) = \frac{i}{\Gamma_0}\sum_{\xi,\xi'}\,\frac{v_{\xi,j}v_{\xi',l}^*}{\lambda_\xi-\lambda_{\xi'}^*}\,s_\xi(0)s_{\xi'}^*(0).
\end{equation}
Inserting this equality into Eq.~\eqref{efficiency}, we readily find
\begin{equation}
\label{efficiency_2}
\eta = \frac{S_{\lambda_0}}{4 F_\mathrm{det}}\,\sum_{j,l}s_{j}(0)K_{jl}s_{l}^*(0),
\end{equation}
where
\begin{equation}
K_{jl} = i\sum_{\xi,\xi'}\,v_{\xi,j}v^*_{\xi',l}\,\frac{E^*_\xi E_{\xi'}}{\lambda_\xi-\lambda_{\xi'}^*},
\end{equation}
and $E_\xi = \sum_m v_{\xi,m}E_m$. Importantly, $K$ is an $N_\mathrm{a}\times N_\mathrm{a}$ Hermitian matrix which depends only on the positions of the atoms and the detection mode, but not on the specific time dependence of the control field (for example, one could apply a $\pi$ pulse that transfers all of the excitation from state $\ket{s}$ to $\ket{e}$ at time $t=0$). The maximum retrieval efficiency is thus given by the initial configuration corresponding to the eigenvector of $K$ with the largest eigenvalue. We should note that while the efficiency $\eta$ of retrieval is independent of the particular profile $\Omega_c(t)$, the shape of the outgoing photon is completely determined by the control field. By time-reversal symmetry, if one wants to store an incoming photon with maximum efficiency, one must first consider its time-reversed shape (i.e., an outgoing photon), find the unique control field $\Omega_c(t)$ that generates such a shape in retrieval, and then apply the time-reversed field $\bar{\Omega}_c(t)$ for storage.

Before proceeding further, we briefly comment on the classical and quantum optical aspects of the calculation presented above. An equation analogous to Eq.~\eqref{efficiency} also applies if the atoms were replaced by classical oscillating dipoles with amplitudes $e_j(t)$. Such an equation corresponds to the projection of the total classical radiated field into a particular spatial mode. The equivalence between classical and quantum equations is not surprising, given that both the propagation of classical and quantum fields are given by Maxwell's equations. In our particular problem of interest, the quantum nature of the field manifests itself by considering field correlations. For example, using Eq.~\eqref{overlap}, one can calculate the second-order correlation function $\braket{\hat{E}_{det}^{\dagger 2}\hat{E}_{det}^2}$. As the atomic state that we consider contains at most one excitation, this correlation function is exactly zero, or perfectly ``anti-bunched," reflecting the fact that only a single photon is emitted.

\section{2D square array} 
While the formalism presented above is general to any ensemble of atoms with known positions, we now apply it to a 2D square array with lattice constant $d$. This case is particularly interesting, as an infinite array of two-level atoms can act as a perfect mirror for incoming light at normal incidence when $d$ is smaller than the atomic resonant wavelength $\lambda_0$ \cite{Abajo2007,Bettles2016b,Shahmoon2017}. Physically, the incoming field guarantees that all the induced atomic dipoles oscillate with the same phase. While such a configuration can in principle emit into various diffraction orders, for $d<\lambda_0$, all of the orders except the one perpendicular to the plane become evanescent, and cannot radiate away energy. With only two channels of emission possible (forward and backward), the scattered field of the array perfectly interferes with an incident resonant photon in the forward direction, leading to complete reflection of light.
Likewise, when an excitation is stored uniformly in the infinite array with $d<\lambda_0$, it is ``selectively radiant'' \cite{Asenjo2017b}, as interference guarantees that the retrieved photon is perfectly emitted into two plane waves normal to the array (we assume that this symmetric emission can be re-combined). While this simple argument hints that a finite array can also be very efficient, what remains is to quantify the error. We thus analyze the retrieval efficiency of an array made of $N_\mathrm{a} = N\times N$ atoms.

As far as the detection mode is concerned, a common mode to project into is a Gaussian beam. There is a technicality, however, since a Gaussian beam is only an approximate~(paraxial) solution to Maxwell's equations. While such an approximation usually suffices, here we anticipate that one can achieve nearly perfect storage and retrieval efficiencies. Consequently, it is not obvious a priori that the small (actual) retrieval errors are not overwhelmed by the error of the paraxial approximation itself. Thus, we consider an exact mode solution for Maxwell's equations (see \ref{ApB} for details), which approaches the Gaussian solution in the limit of large beam waist $w_0$.

%%%%%%%%%%%%%%%%%%%%%%%%%%%%%%%%%%%%%%%%%%%%%%%%%%%%%%%%%%%%%%%%%%%%%%%%%%%%%%%%%%%%
\begin{figure}[t]
\includegraphics[width=13cm,right]{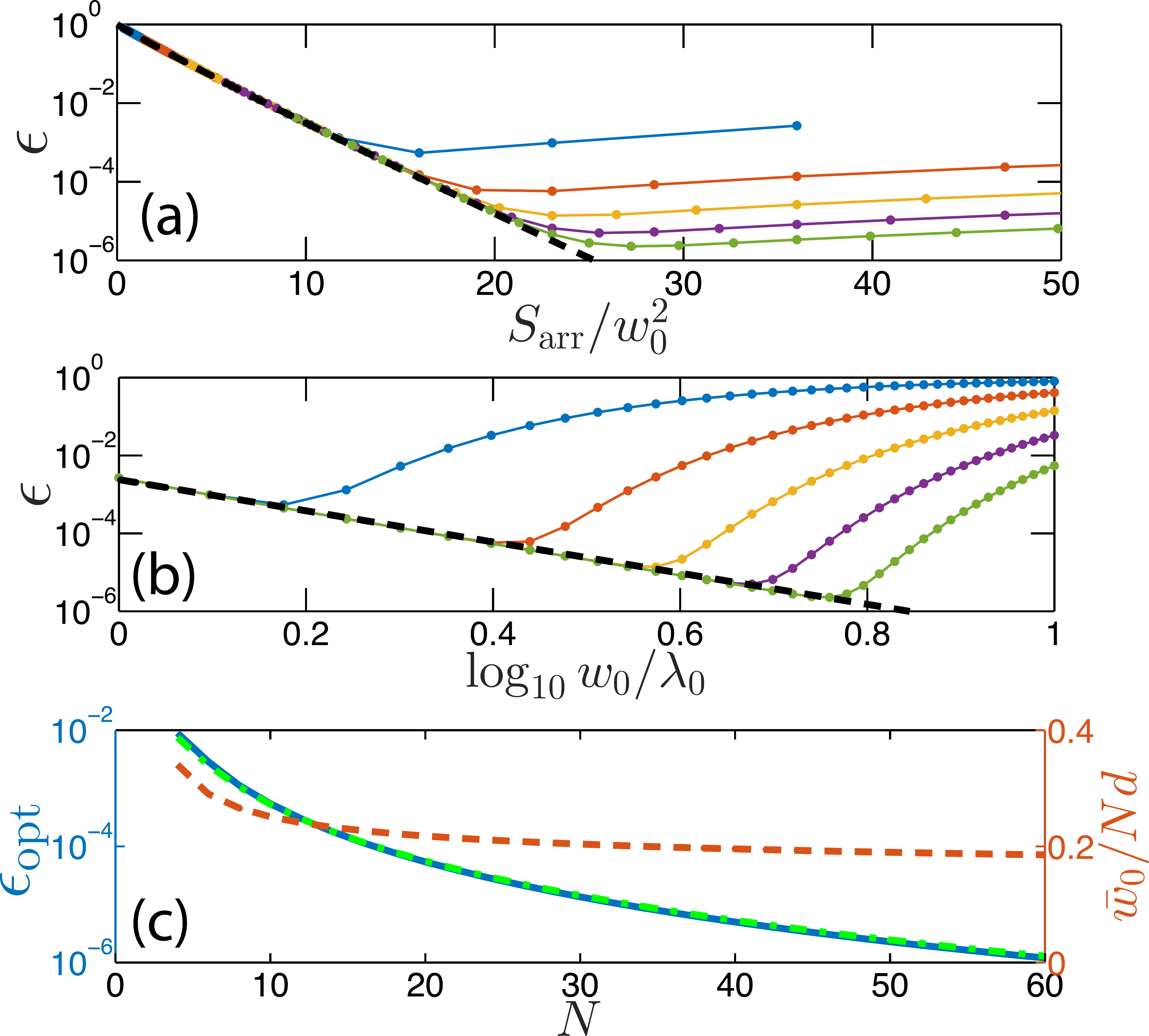}
\caption{Minimum retrieval error $\epsilon = 1-\eta$ from a square array of atoms into a Gaussian-like detection mode, as functions of (a) $S_\mathrm{arr}/w_0^2$, with $S_\mathrm{arr} = (Nd)^2$ being the array area, and (b) $\log_{10}w_0/\lambda_0$ for $d = 0.6\lambda_0$ and $N = 10, 20, 30, 40, 50$ (blue, red, yellow, violet, green, respectively). The black dashed lines in (a) and (b) correspond respectively to $\epsilon = 1-\text{Erf}^2(Nd/\sqrt{2}w_0)$ and $\epsilon = C(\lambda_0/w_0)^4$. (c) Left axis: value of $\epsilon_\mathrm{opt}$ for a beam waist $\bar{w}_0$ obtianed from numerical optimization (blue continuous line), and the approximate analytical error of Eq.~\eqref{approx_err} (green dot-dashed line). Right axis: ratio between the optimal beam waist $\bar{w}_0$ and the linear dimension of the array $Nd$, as a function of $N$ (red dashed line). }
\label{fig:efficiency}
\end{figure}
%%%%%%%%%%%%%%%%%%%%%%%%%%%%%%%%%%%%%%%%%%%%%%%%%%%%%%%%%%%%%%%%%%%%%%%%%%%%%%%%%%%%

Before presenting the numerics, one can already intuitively argue the fundamental sources of error associated with a finite array by considering the reflectance problem. If the beam waist $w_0$ is too large with respect to the array dimensions, then part of the incoming light will not see the atoms and will be transmitted or scattered in other directions by the edges of the array. If $w_0$ is too small, the incoming mode contains a broad range of wavevectors with different propagation directions. Since different angles have maximum reflectance at different detunings relative to the bare transition frequency $\omega_{eg}$ \cite{Shahmoon2017}, the overall reflectance for a near-monochromatic photon will be reduced. For a given array, an optimal beam waist thus maximizes the reflectance of an incoming photon (at optimal detuning). The situation is analogous for the retrieval problem, where the optimization over the photon frequency is replaced by an optimization over the initial spatial distribution of the collective $s$-excitation.

To check this behavior, we numerically calculate the minimum retrieval error $\epsilon=1-\eta$ varying the beam waist $w_0$, for several different atom numbers. In Fig.~\ref{fig:efficiency}(a), the error is plotted as a function of the ratio between the array area $S_\mathrm{arr}=d^2N_\mathrm{a}$ and $w_0^2$. Here, we have taken the retrieval mode to consist of a symmetric superposition of Gaussian beams emitted in opposite directions from the array, with the view that these beams can in principle be recombined. For concreteness, we consider a lattice constant of $d=0.6\lambda_0$, although other choices $d<\lambda_0$ do not affect the general scalings. As $S_\mathrm{arr}/w_0^2$ grows, the error initially scales as $\epsilon\sim 1-\text{Erf}^2(Nd/\sqrt{2}w_0)$ (illustrated by the dashed curve), where $\text{Erf}(z)$ is the error function. Physically, this error corresponds to the fraction of the energy carried by the Gaussian beam beyond the array boundaries. In Fig.~\ref{fig:efficiency}(b) we plot (in log-log scale) $\epsilon$ as a function of the ratio between $w_0$ and $\lambda_0$ (for values larger than one), again for different array sizes. Up to a point where the beam waist becomes comparable with the array dimension, the error scales roughly as $\epsilon \sim (\lambda_0/w_0)^4$ (dashed line). This error physically arises from the range of wavevector components that make up the detection mode, which is inversely proportional to $w_0$. An analysis of the reflectance of a beam of finite waist from an infinite array in fact shows the same scaling, when considering the fraction of light that is not reflected. Overall we have that the minimum error can be approximated by the expression
\begin{equation}
\label{fit}
\epsilon(N,d,w_0) \approx C(d)(\lambda_0/w_0)^{4} + 1 - \text{Erf}^2(Nd/\sqrt{2}w_0).
\end{equation}
The constant $C$ can be obtained by fitting the error: for $d = 0.6\lambda_0$ we find $C \approx 2.4\cdot 10^{-3}$.

One can use Eq.~\eqref{fit} to find the optimal beam waist. After optimizing $w_0$ we find that the leading term for the error is given by
\begin{equation}
\label{approx_err}
\epsilon_\mathrm{opt} \approx (\log N_\mathrm{a})^2/(4N_\mathrm{a}^2).
\end{equation}
In Fig.~\ref{fig:efficiency}(c) this approximate expression for the minimum retrieval error is compared with the value obtained by numerical optimization. The associated optimum beam waist for the retrieval mode is also plotted for completeness. Interestingly, even a $4\times 4$ array of atoms can in principle already enable a storage/retrieval efficiency of above 99\%. In comparison, an optical depth of nearly $D\sim 600$ is needed to obtain the same error in a conventional ensemble~\cite{Gorshkov2007}. In the case where the beam waist does not significantly diverge over the length of the ensemble, the optical depth is given by $D\sim S_{\lambda_0} N_{a}/w_0^2$. For cold atoms, an atom number on the order of $N_a\sim 10^6$-$10^7$ might be required to achieve a value of $D\sim 600$.

\section{Relevant Imperfections}
\subsection{Analysis of disorder}

\begin{figure}[t]
\includegraphics[width=13cm,angle=0,clip,right]{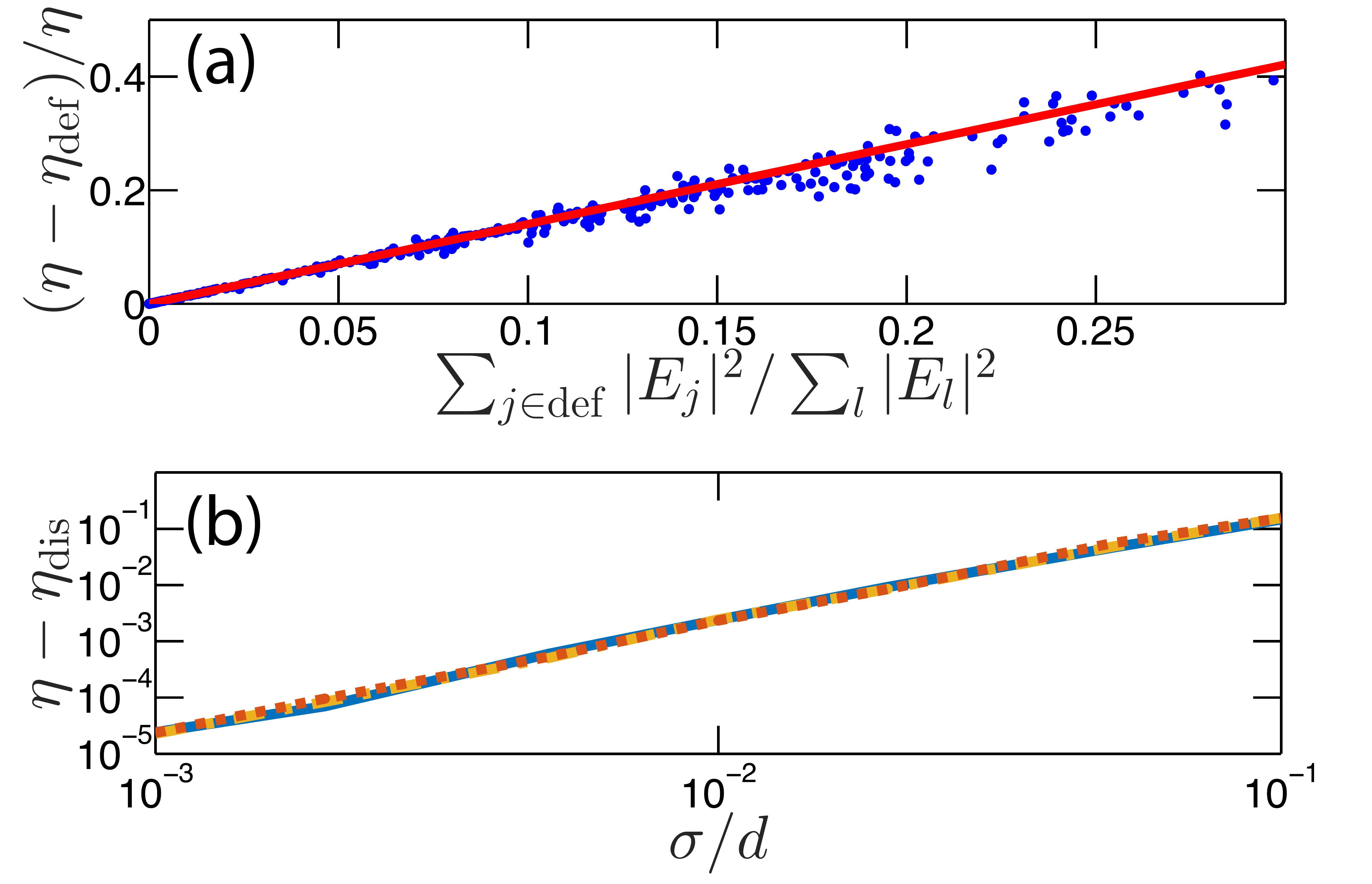}
\caption{(a) Relative difference between the perfect array efficiency and efficiency of an array with ``holes" as a function of $\sum_{j\in \text{def}}|E_j|^2/\sum_l|E_l|^2$. Each dot represents a random defect configuration of a $10 \times 10$ array with fixed $d=0.6\lambda_0$ and $w_0=1.5\lambda_0$. For each number of holes from 1 to 20, 100 configurations are considered (only 20 are represented for visual clarity). The red line is a linear fit. (b) Difference between the optimized maximum retrieval efficiency $\eta$ and the mean retrieval efficiency $\eta_\mathrm{dis}$ obtained using the same initial conditions and beam waist but with position disorder $\sigma$ in the atomic positions (log-log scale). The different colors correspond to $N = 6,10,20$ (blue, red, yellow, respectively), with $d = 0.6\lambda_0$. For each value of $\sigma$, 100 random configurations are considered.}
\label{fig:eff_disorder}
\end{figure}

In this section, we analyze the effects of various types of disorder in the array. One useful attribute of our efficiency calculation is that it enables different spatial configurations to be studied. Thus, we can easily include imperfections such as the absence of atoms (``holes") in the array, or classical position disorder. We first examine the case of some number $N_\mathrm{def}$ of holes in the array. Intuitively, one expects that the relative decrease in efficiency, $(\eta-\eta_\mathrm{def})/\eta$, will be proportional to the ratio between the intensity of the detection mode hitting the empty sites, to the total intensity over the array. Here, $\eta_\mathrm{def}$ and $\eta$ denote the maximum retrieval efficiency with and without the holes, respectively, with the beam waist $w_0$ chosen to optimize $\eta$. In Fig.~\ref{fig:eff_disorder}(a) we plot the relative loss as a function of $\sum_{j\in \text{def}}|E_j|^2/\sum_l|E_l|^2$, where the sums of the field intensities in the numerator and denominator run over sites of holes and all sites, respectively, sampling over 100 random configurations for different densities of holes ($N_\mathrm{def}/N_\mathrm{a}$ up to 20\%). One sees a clear statistical relation of the form
\begin{equation}
\label{defect}
\eta_{\mathrm{def},\{j\}} \sim \eta\bigg(1 - \alpha\frac{\sum_{j\in \text{def}}|E_j|^2}{\sum_l|E_l|^2}\bigg).
\end{equation}
The constant of proportionality $\alpha$ in Eq.~\eqref{defect} depends only on $d$ and is about $\alpha\approx 1.25$ for $d = 0.6\lambda_0$. While here we have optimized the initial spin wave for each random configuration, which would be applicable if an experiment could resolve the positions of the holes in a single shot \cite{Endres2016}, we expect a similar scaling even if the positions of holes are unknown.

%%%%%%%%%%%%%%%%%%%%%%%%%%%%%%%%%%%%%%%%%%%%%%%%%%%%%%%%%%%%%%%%%%%%%%%%%%%%%%%%%%%%
%%%%%%%%%%%%%%%%%%%%%%%%%%%%%%%%%%%%%%%%%%%%%%%%%%%%%%%%%%%%%%%%%%%%%%%%%%%%%%%%%%%%

Classical disorder for the atomic positions consists in having the atoms displaced by random amounts $\delta_j = (\delta_{x,j},\delta_{y,j})$ from their position in the perfect lattice. It is shown in Ref.~\cite{Shahmoon2017} for the case of reflectance of an infinite array that, when the $\delta$'s are extracted from a Gaussian distribution with standard deviation $\sigma$, then the decrease in reflectance introduced by the disorder scales as $\sigma^2/d^2$. We find numerically the same result for the retrieval error of the finite array. In particular, in Fig.~\ref{fig:eff_disorder}(b) the error introduced by disorder is plotted as a function of $\sigma$ for different array dimensions and fixed lattice constant. This error is defined as the difference between the optimized maximum retrieval efficiency $\eta$ of a perfect lattice, and the mean retrieval efficiency $\eta_\mathrm{dis}$ (sampled over many configurations) with the same initial atomic wave function and beam waist but with disorder in the atomic positions.

\subsection{Finite detection time}

When calculating the retrieval efficiency, given by Eq.~\eqref{efficiency}, we have implicitly assumed that the detection time is infinite, such that all the energy emitted into the detection mode is collected. Practically, it might also be relevant to consider the retrieval efficiency given a finite time window $0<t<T_{\mathrm{d}}$ for photon collection, such as if an experiment has other limiting time scales (i.e., atom trapping time, required fast readout, etc.).

The efficiency detected for an arbitrary detection time window $T_\mathrm{d}$ is given by
\begin{equation}
\eta_{T_\mathrm{d}} = \frac{S_{\lambda_0}\Gamma_0}{4F_\mathrm{det}}\,\sum_{j,l}\,E^*_jE_l\,\int_0^{T_\mathrm{d}} dt\,e_j(t)e_l^*(t),
\end{equation}
where $e_j(t)$ is obtained by integrating Eqs.~\eqref{eom}-\eqref{eom_2}. In general the temporal profile of the emitted field depends on the control field amplitude $\Omega_c(t)$ and detuning $\Delta$. If one wants to achieve a high efficiency in the shortest time, then the optimal strategy is to essentially use the control field to apply an instantaneous $\pi$-pulse at $t=0$, thus instantly transferring the excitation stored in the metastable state $\ket{s}$ to the rapidly emitting excited state $\ket{e}$. In an array, this collective excitation in $\ket{e}$ will emit a photon at a rate $\sim\Gamma_0$ comparable to the single-atom emission rate, ensuring that the errors due to finite time window $T_\mathrm{d}$ become very small once $T_\mathrm{d}$ is on the order of a few $\sim\Gamma_0^{-1}$.

\begin{figure}[t]
\includegraphics[width=14cm,angle=0,clip,right]{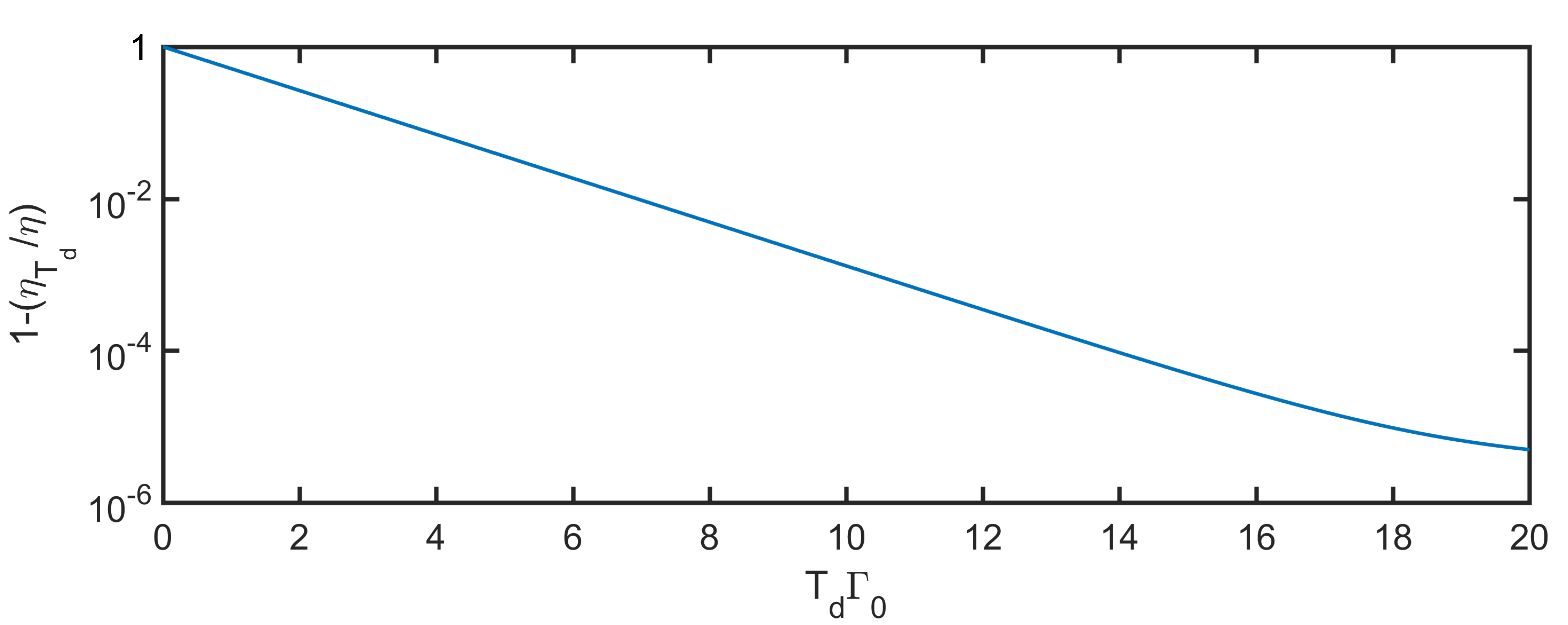}
\caption{Relative reduction in the retrieval efficiency $1 - \eta_{T_\mathrm{d}}/\eta$ as function of the detection time window $T_\mathrm{d}$ (lin-log scale) for an array of $10\times 10$ atoms with $d=0.6\lambda_0$ and optimal beam waist.}
\label{fig:finite_detection}
\end{figure}

In Fig.~\ref{fig:finite_detection}, we plot the relative error $1 - \eta_{T_\mathrm{d}}/\eta$ due to the finite detection time, where $\eta$ is the detection efficiency for an infinite time window, for an array of $10\times 10$ atoms with $d=0.6\lambda_0$ and optimal beam waist. We notice that for a detection time ${T_\mathrm{d}}\sim 10/\Gamma_0$ the error is of the order of $10^{-3}$. The possibility of having a good retrieval efficiency even for a short detection time is a consequence of the fact that, while the array can support highly subradiant states~\cite{Bettles2015,Asenjo2017b,Facchinetti2016,Bettles2016b,Shahmoon2017}, they form a negligible component of the \textit{optimized} spin wave for storage and retrieval. This makes intuitive sense, as to interface with light efficiently, one should use radiant or ``selectively radiant" atomic excitations rather than states that decouple from light.

%%%%%%%%%%%%%%%%%%%%%%%%%%%%%%%%%%%%%%%%%%%%%%%%%%%%%%%%%%%%%%%%%%%%%%%%%%%%%%%%%%%% 
%%%%%%%%%%%%%%%%%%%%%%%%%%%%%%%%%%%%%%%%%%%%%%%%%%%%%%%%%%%%%%%%%%%%%%%%%%%%%%%%%%%%

\section{Conclusions} 
In summary, we have introduced a prescription to calculate the maximum storage and retrieval efficiency of a quantum memory, which fully accounts for re-scattering and interference of light emission in all directions. Our approach is in principle applicable to any system where the positions of the emitters are known (or can be reasonably modelled, such as assigning random positions) and the spatial and spectral response of the dielectric environment (i.e., the Green's function) is also known \cite{Julsgaard2004,Choi2008,Liu2001,Lukin2001b,Vetsch2010a, KGB05, Asenjo2017b, Hood2016,Asenjo2017,Lalumiere2013,LFL13,maxwell, sipahigil}. As one particular application, we have shown an improved scaling of errors for atoms in free space, compared to the result predicted by the one-dimensional Maxwell-Bloch equations. We speculate that it is possible to obtain an exponential reduction of errors versus atom number in free space, by using arrays that are not completely periodic. The question of how to tailor the spatial positions will be left to future work.

More broadly, we expect that a significantly improved storage efficiency is possible whenever the excited state emission is largely radiative and coherent, which includes not only atoms but solid-state emitters with large zero-phonon line and Fourier-limited linewidths~\cite{sipahigil}. Techniques to reversibly map between photonic and atomic excitations in arrays should find a variety of exciting applications. For example, it would allow for photonic quantum gates, given some form of spin interactions in the array (such as between Rydberg levels \cite{Labuhn2016}), or would allow for exotic spin states (like subradiant \cite{Zoubi2010,Bettles2015,Bettles2016,Sutherland2016,Hebenstreit2017,Asenjo2017b} or topological excitations \cite{Perczel2017,Bettles2017}) to be detected optically. It would also be interesting to investigate whether the spin state itself could be engineered to produce a useful non-classical state of outgoing light. More broadly, the ability to formally map atom-light interactions to a long-range open spin model could provide new insights into quantum optical phenomena with atomic systems.

\appendix
\section{Green's function expansion in plane and evanescent waves}\label{ApA}

Here we derive Eq.~\eqref{overlap} of the main text by using an expansion of the Green's function in terms of plane and evanescent waves. The Green's function Eq.~\eqref{Greens} can be written in the angular spectrum representation, \emph{i.e.}~as an integral over $k_x$ and $k_y$ in Fourier space, as \cite{Novotny2006} 
\begin{equation}
\label{G_exp}
\mathbf{G}_0^\pm(\mathbf{r}, \mathbf{r}', \omega_{eg}) = \frac{i}{8\pi^2k^2_0}\int^{\infty}_{-\infty}\int^{\infty}_{-\infty}\frac{dk_xdk_y}{k_z}\,\mathbf{Q}^{\pm}\,e^{ik_x(x-x')+ik_y(y-y')\pm ik_z(z-z')},
\end{equation}
where
\begin{equation}
\mathbf{Q}^{\pm} =
\begin{pmatrix}
k_0^2 - k_x^2 & -k_xk_y & \mp k_xk_z \\
-k_xk_y & k_0^2 - k_y^2 & \mp k_yk_z \\
\mp k_xk_z & \mp k_yk_z & k_0^2 - k_z^2
\end{pmatrix}
\end{equation}
and the $\pm$ denoting the sign of $z-z'$. We can separate the integral in Eq.~\eqref{G_exp} into two separate integrals: for values of $k_x$, $k_y$ lying inside and outside the disk defined by $k_x^2 + k_y ^2 = k_0^2$. This decomposition separates the plane waves from evanescent waves, \emph{i.e.}, we can write $\mathbf{G}^\pm(\mathbf{r}, \mathbf{r}', \omega_{eg}) = \mathbf{G}^\pm_\mathrm{pl}(\mathbf{r}, \mathbf{r}', \omega_{eg}) + \mathbf{G}^\pm_\mathrm{ev}(\mathbf{r}, \mathbf{r}', \omega_{eg})$, where
\begin{equation}
\mathbf{G}^\pm_\mathrm{pl}(\mathbf{r}, \mathbf{r}', \omega_{eg}) = \frac{i}{8\pi^2k^2_0}\int_{k_x^2+k_y^2\leq k_0^2}\frac{dk_xdk_y}{k_z}\,
\mathbf{Q}^{\pm}
\,e^{ik_x(x-x')+ik_y(y-y')\pm ik_z(z-z')},
\end{equation}
with $k_z = \sqrt{k_0^2-k_x^2-k_y^2}$,
and 
\begin{equation}
\mathbf{G}^\pm_\mathrm{ev}(\mathbf{r}, \mathbf{r}', \omega_{eg}) = \frac{1}{8\pi^2k^2_0}\int_{k_x^2+k_y^2> k_0^2}\frac{dk_xdk_y}{k_z}\,
\mathbf{Q}^{\pm}
\,e^{ik_x(x-x')+ik_y(y-y')\pm ik_z(z-z')},
\end{equation}
with $k_z = i\sqrt{k_x^2+k_y^2-k_0^2}$.

The integral in the plane waves part can be rewritten in polar coordinates using $\mathbf{k}_0 = k_0(\sin\theta\cos\phi,\sin\theta\sin\phi,\cos\theta)$, obtaining
\begin{equation}
\label{G_plane}
\mathbf{G}^\pm_\mathrm{pl}(\mathbf{r}, \mathbf{r}', \omega_{eg}) = \frac{i}{8\pi^2k_0}\int^{2\pi}_{0}d\phi\int^{\pi/2}_0d\theta\sin\theta\,\mathbf{Q}^{\pm}\,e^{ik_0(\sin\theta\cos\phi(x-x')+\sin\theta\sin\phi(y-y')\pm\cos\theta(z-z'))}.
\end{equation}
It can be shown easily that, introducing the polarization vectors 
\begin{eqnarray}
\hat{e}^1_{\mathbf{k}_0}&=&(\sin\phi,-\cos\phi,0), \\
\hat{e}^2_{\mathbf{k}_0}&=&(\cos\theta\cos\phi,\cos\theta\sin\phi,-\sin\theta),
\end{eqnarray}
orthogonal to $\mathbf{k}_0$ and between them, $\mathbf{G}^+_\mathrm{pl}(\mathbf{r}, \mathbf{r}', \omega_{eg})$ can be expressed as 
\begin{equation}
\label{G_p_pw}
\mathbf{G}^+_\mathrm{pl}(\mathbf{r}, \mathbf{r}', \omega_{eg}) = \frac{ik_0}{8\pi^2}\sum_{\alpha}\int^{2\pi}_{0}d\phi\int^{\pi/2}_0d\theta\sin\theta\,\mathbf{u}^*_{k_0,\theta,\phi,\alpha}(\mathbf{r})\mathbf{u}_{k_0,\theta,\phi,\alpha}(\mathbf{r}'),
\end{equation}
where we have defined a plane wave basis
\begin{equation}
\mathbf{u}_{k_0,\theta,\phi,\alpha}(\mathbf{r}) = \hat{e}^\alpha_{\mathbf{k}_0}\,e^{-ik_0(\sin\theta\cos\phi x+\sin\theta\sin\phi y+\cos\theta z)},
\end{equation}
with the normalization
\begin{equation}
\label{normalization}
\int_{z=\text{const}} d^2\mathbf{r}\,\mathbf{u}^*_{k_0,\theta,\phi,\alpha}(\mathbf{r})\cdot\mathbf{u}_{k_0,\theta',\phi',\beta}(\mathbf{r}) = \frac{(2\pi)^2}{k_0^{2}\sin\theta}\delta(\theta-\theta')\delta(\phi-\phi')\delta_{\alpha\beta}.
\end{equation}
Similarly one has 
\begin{equation}
\label{G_m_pw}
\mathbf{G}^-_\mathrm{pl}(\mathbf{r}, \mathbf{r}', \omega_{eg}) = \frac{ik_0}{8\pi^2}\sum_{\alpha}\int^{2\pi}_{0}d\phi\int^{\pi}_{\pi/2}d\theta\sin\theta\,\mathbf{u}^*_{k_0,\theta,\phi,\alpha}(\mathbf{r})\mathbf{u}_{k_0,\theta,\phi,\alpha}(\mathbf{r}').
\end{equation}
%where we have defined $\sum_{\mathbf{k}_0} = (k_0/2\pi)^2\int^{2\pi}_0 d\phi\int^{\pi}_0 d\theta \sin\theta \sum_{\alpha}$.

An analogous expression can be found for the evanescent wave part. Here it is convenient to define the vector $\tilde{\mathbf{k}}_0 = k_0(\cosh\xi\cos\phi,\cosh\xi\sin\phi,i\sinh\xi)$: 
\begin{equation}
\mathbf{G}_\mathrm{ev}^{\pm}(\mathbf{r}, \mathbf{r}', \omega_{eg}) = \frac{1}{8\pi^2k_0}\int^{2\pi}_{0}d\phi\int^{\infty}_0d\xi\sinh\xi\,\mathbf{Q}^{\pm}\,e^{k_0(i\cosh\xi\cos\phi(x-x')+i\cosh\xi\sin\phi(y-y')\mp\sinh\xi(z-z'))}.
\end{equation}
With the polarization vectors defined by
\begin{eqnarray}
\hat{e}^1_{\tilde{\mathbf{k}}_0}&=&(\sin\phi,-\cos\phi,0), \\
\hat{e}^2_{\tilde{\mathbf{k}}_0}&=&(-i\sinh\xi\cos\phi,-i\sinh\xi\sin\phi,-\cosh\xi)
\end{eqnarray}
orthogonal to $\tilde{\mathbf{k}}_0$ and between them, one can indeed write 
\begin{equation}
\mathbf{G}^+_\mathrm{ev}(\mathbf{r}, \mathbf{r}', \omega_{eg}) = \frac{k_0}{8\pi^2}\sum_{\alpha}\int^{2\pi}_{0}d\phi\int^{\infty}_0d\xi\sinh\xi\,\tilde{\mathbf{u}}^*_{k_0,\xi,\phi,\alpha}(\mathbf{r})\tilde{\mathbf{u}}_{k_0,\xi,\phi,\alpha}(\mathbf{r}'),
\end{equation}
where 
\begin{equation}
\tilde{\mathbf{u}}_{k_0,\xi,\phi,\alpha}(\mathbf{r}) = \hat{e}^\alpha_{\tilde{\mathbf{k}}_0}\,e^{-k_0(i\cosh\xi\cos\phi x+i\cosh\xi\sin\phi y-\sinh\xi z)}.
\end{equation}
Similarly one has
\begin{equation}
\mathbf{G}^-_\mathrm{ev}(\mathbf{r}, \mathbf{r}', \omega_{eg}) = \frac{k_0}{8\pi^2}\sum_{\alpha}\int^{2\pi}_{0}d\phi\int^{0}_{-\infty}d\xi\sinh\xi\,\tilde{\mathbf{u}}^*_{k_0,\xi,\phi,\alpha}(\mathbf{r})\tilde{\mathbf{u}}_{k_0,\xi,\phi,\alpha}(\mathbf{r}').
\end{equation}

Now let's consider a detection mode that does not contain evanescent components for simplicity, so that it can be expanded just in terms of monochromatic plane waves as 
\begin{equation}
\mathbf{E}_\mathrm{det}(\mathbf{r}) = \frac{1}{(2\pi)^2}\sum_{\alpha}\int^{2\pi}_{0}d\phi\int^{\pi}_0d\theta\sin\theta\,c_{k_0,\theta,\phi,\alpha}\mathbf{u}_{k_0,\theta,\phi,\alpha}(\mathbf{r}).
\end{equation}
The overlap between this mode and the field generated by a dipole is
\begin{multline}
\braket{\mathbf{E}_\mathrm{det}|\mathbf{E}_\mathrm{out}} = \int_{z=\mathrm{const}} d^2\mathbf{r}\,\mathbf{E}_\mathrm{det}^*(\mathbf{r})\cdot \mathbf{E}_\mathrm{out}(\mathbf{r}) = \\ = \frac{id_{eg}k^3_0}{2\epsilon_0(2\pi)^4}\,\int_{z=\mathrm{const}} d^2\mathbf{r}\,\sum_{\alpha,\beta}\int^{2\pi}_{0}d\phi\int^{\pi}_0d\theta\sin\theta\int^{2\pi}_{0}d\phi'\int^{\pi}_0d\theta'\sin\theta'\,\times\\
\times\,c_{k_0,\theta,\phi,\alpha}\mathbf{u}_{k_0,\theta,\phi,\alpha}(\mathbf{r})\,\mathbf{u}^*_{k_0,\theta',\phi',\beta}(\mathbf{r}_\mathrm{d})\mathbf{u}_{k_0,\theta',\phi',\beta}(\mathbf{r})\cdot \hat{\mathbf{d}}\sigma^{ge} \\ = \frac{id_{eg}k_0}{2\epsilon_0(2\pi)^2}\,\sum_{\alpha}\int^{2\pi}_{0}d\phi\int^{\pi}_0d\theta\sin\theta\,c_{k_0,\theta,\phi,\alpha}\mathbf{u}^*_{k_0,\theta',\phi',\beta}(\mathbf{r}_\mathrm{d})\cdot \hat{\mathbf{d}}\sigma^{ge} = \frac{i d_{eg}k_0}{2\epsilon_0 }\mathbf{E}_\mathrm{det}^*(\mathbf{r}_\mathrm{d})\cdot \hat{\mathbf{d}}\sigma^{ge}.
\end{multline}  
where we have used Eq.~\eqref{output} (without input field) to express the field generated by the dipole through the Green's function and Eqs.~\eqref{G_p_pw} and \eqref{G_m_pw} for the Green's function decomposition. Adding the input field and normalizing the detection mode we finally obtain Eq.~\eqref{overlap} of the main text.

\section{Gaussian detection mode}\label{ApB}

Here we present the detection mode which we have chosen to study the retrieval efficiency of the 2D array. We choose a solution oscillating with frequency $e^{-i\omega_{eg} t}$, and where the $x$-component of the electric field in wavevector space is given by $E_x(k_x,k_y)\propto e^{-(k_x^2+k_y^2)w_0^2/4}\Theta(k_0^2-k_x^2-k_y^2)$, where $\Theta(x)$ is the Heaviside step function. That is, $E_x$ has a Gaussian distribution for $k_x^2+k_y^2\leq k_0^2$ while it is zero for $k_x^2+k_y^2>k_0^2$, such that the field does not contain evanescent components. In the $y$ direction, we take the field to be identically zero. The value of the $z$-component is then determined by Maxwell's equations \cite{Chen2002}. The real space profile of this mode can be obtained by Fourier transformation:
\begin{equation}
\label{as_2_real_x}
E_\mathrm{det}^x(\mathbf{r}) = E_0\,\int^{1}_0 db\,b\,e^{-b^2k^2_0w_0^2/4}\,e^{i k_0z\sqrt{1-b^2}}\,J_0(bk_0\rho),
\end{equation}
and
\begin{equation}
\label{as_2_real_z}
E_\mathrm{det}^z(\mathbf{r}) = -i E_0\,\frac{x}{\rho}\int^{1}_0 db\,\frac{b^2}{\sqrt{1-b^2}}\,e^{-b^2k_0^2w_0^2/4}\,e^{i k_0z\sqrt{1-b^2}}\,J_1(bk_0\rho),
\end{equation}
where $(\rho,z)$ are the cylindrical coordinates for $\mathbf{r}$, while $J_0$ and $J_1$ are Bessel functions. If evanescent components were included, the field in real space would identically consist of a Gaussian in the $z=0$ focal plane with beam waist $w_0$. The step function in wavevector space enforces in real space a diffraction limit, and distorts the beam to prevent a focal spot smaller than $\sim\lambda_0$. For large $w_0$ the mode tends to the paraxial solution, \emph{i.e.}~$E_\mathrm{det}^z$ vanishes and $E_\mathrm{det}^x$ assumes the form of a fundamental Laguerre-Gauss mode \cite{Novotny2006}.

\section{Spin model for isotropic atoms}\label{ApC}

In the main text we have introduced a formalism to calculate the retrieval efficiency of an atomic ensemble of three-level atoms, with an excitation initially stored in a metastable state $\ket{s}$ coupled to the excited state $\ket{e}$ by a classical control field. Instead of a single excited state, a more realistic minimal model of an atom consists of three excited states $\ket{e_\alpha}$, where $\alpha=x,y,z$ denotes the three possible orientations of the dipole transition $\hat{\mathbf{d}}$.
The effective Hamiltonian \eqref{eff_Ham} generalizes to
\begin{equation}
\label{H_deg}
H_\mathrm{eff} = H_\mathrm{in} - \mu_0d_{eg}^2\omega_{eg}^2\sum_{j,l}\sum_{\alpha\beta}G_{\alpha\beta}(\mathbf{r}_j,\mathbf{r}_l,\omega_{eg})\,\sigma^{eg}_{\alpha,j}\sigma^{ge}_{\beta,l},
\end{equation}
where the sum over $\alpha$ and $\beta$ are over $x,y,z$. Here, $\sigma^{ge}_{\beta,l}=\ket{g}_l\bra{e_\beta}_l$ is the lowering operator on atom $l$, which takes the excited state $\ket{e_\beta}$ to the ground state $\ket{g}$. It should be noted that in general, transitions with different orientations can mix together (\emph{e.g.}, one atom could decay from $\ket{e_y}$ and excite another atom from the ground state to $\ket{e_x}$), as a photon emitted from a given dipole orientation does not have the same global polarization everywhere in space.

\begin{figure}[t]
\includegraphics[width=14cm,angle=0,clip,right]{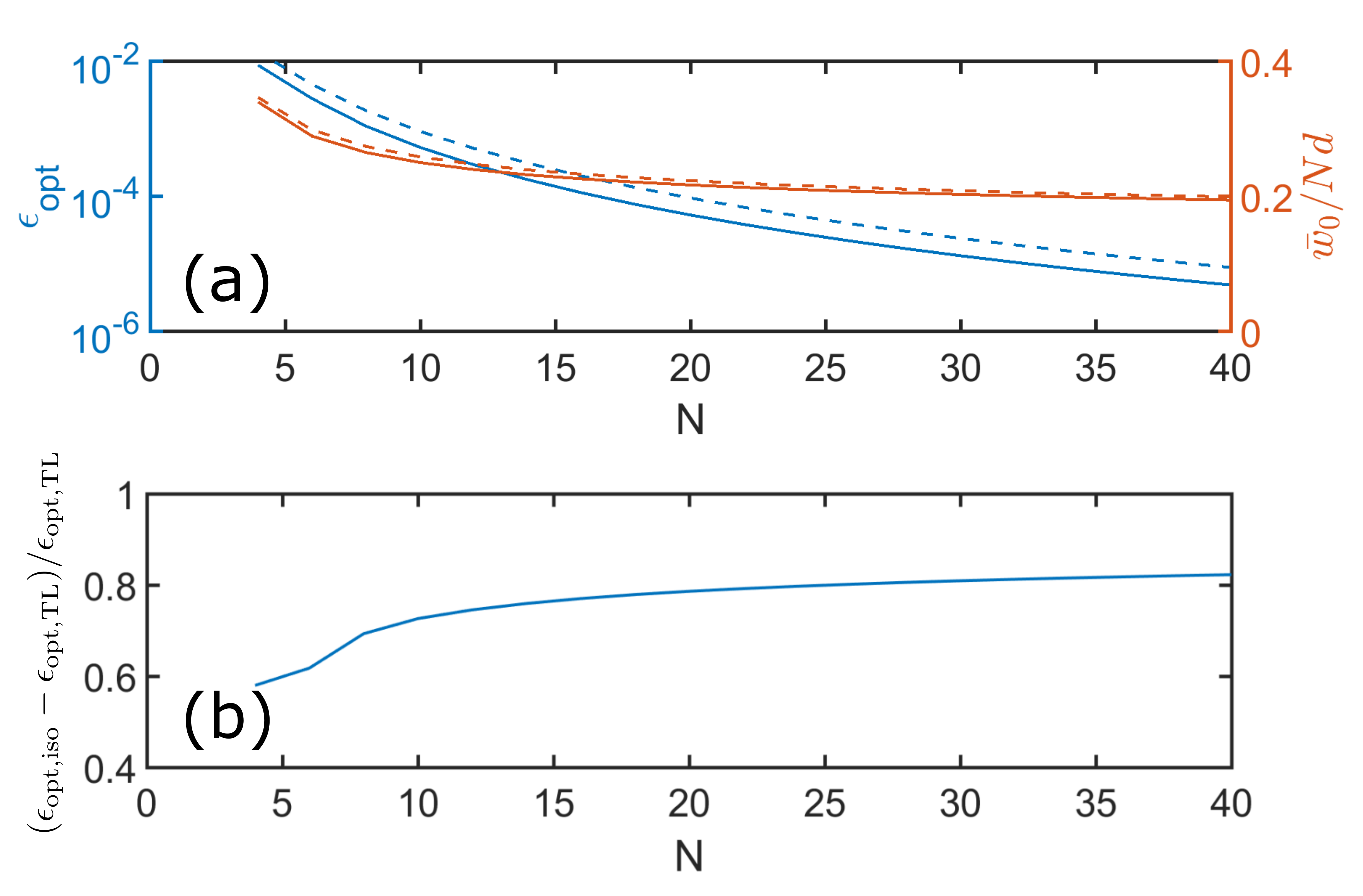}
\caption{(a) Comparison between the optimal retrieval error $\epsilon_\mathrm{opt}$ (left axis, blue lines) and the corresponding optimal beam waist $\bar{w}_0$ (right axis, red lines) for the case of a single excited state discussed in the main text (continuous lines) and the case of three excited states (dashed lines), as functions of the linear array dimension $N$. (b) Relative difference $(\epsilon_\mathrm{opt,iso}-\epsilon_\mathrm{opt,TL})/\epsilon_\mathrm{opt,TL}$ between the retrieval errors of the isotropic and two-level atomic structures plotted in (a).}
\label{fig:eff_iso}
\end{figure}

In the case in which the state $\ket{s}$ is coupled only to one of the three excited states, for concreteness $\ket{e_x}$, it is straightforward to generalize the main result of the paper. Eq.~\eqref{efficiency_2} indeed keeps the same form, but with the matix $K$ generalized to 
\begin{equation}
K_{jl} = i\sum_{\xi,\xi'}\,v_{\xi;j,x}v^*_{\xi';l,x}\,\frac{E^*_\xi E_{\xi'}}{\lambda_\xi-\lambda_{\xi'}^*},
\end{equation}
where $E_\xi = \sum_m v_{\xi,m}{E^x_\mathrm{det}}(\mathbf{r}_m)$ and the sum over the index $\xi$ of the eigenvectors has $3N_\mathrm{a}$ values. In Fig.~\ref{fig:eff_iso}(a) we compare the minimum retrieval error for an $N\times N$ square array of atoms versus $N$, for the cases of a single excited state and for the three-fold degenerate excited states. We notice that, while the scaling of the error remains the same, a small reduction of the efficiency is observable in the isotropic case, a consequence of the fact that light polarized along $y$ can be emitted from atoms in the state $\ket{e_x}$ with a reduction of the overlap between the output mode and detection mode. The increase of the error is better quantified in Fig.~\ref{fig:eff_iso}(b) where the relative difference is plotted. We observe that for the range of array sizes considered here the error increases between 50\% and 90\%. The value of the optimal beam waist is instead not particularly affected, as expected.

%%%%%%%%%%%%%%%%%%%%%%%%%%%%%%%%%%%%%%%%%%%%%%%%%%%%%%%%%%%%%%%%%%%%%%%%%%%%%%%%%%%% 
%%%%%%%%%%%%%%%%%%%%%%%%%%%%%%%%%%%%%%%%%%%%%%%%%%%%%%%%%%%%%%%%%%%%%%%%%%%%%%%%%%%% 

\ack
We are grateful to H.~J. Kimble, S. Yelin, M.~D. Lukin, E. Shahmoon, Y. Wang, and M. Gullans for stimulating discussions. M.T.M. was supported by the ``la Caixa-Severo Ochoa'' PhD Fellowship. A.A.-G. was supported by an IQIM postdoctoral fellowship and the Global Marie Curie Fellowship LANTERN. A.V.G. acknowledges support from ARL CDQI, ARO MURI, NSF QIS, AFOSR, NSF PFC at JQI, and ARO. D.E.C. acknowledges support from Fundacio Privada Cellex, Spanish MINECO Severo Ochoa Program SEV-2015-0522, MINECO Plan Nacional Grant CANS, CERCA Programme/Generalitat de Catalunya, and ERC Starting Grant FOQAL.

\section*{References}
\bibliographystyle{unsrt}
\bibliography{quant_mem_ref}

\begin{thebibliography}{10}

\bibitem{Hammerer2010a}
Klemens Hammerer, Anders~S. S\o{}rensen, and Eugene~S. Polzik.
\newblock Quantum interface between light and atomic ensembles.
\newblock {\em Rev. Mod. Phys.}, 82(2):1041--1093, 2010.

\bibitem{Julsgaard2004}
Brian Julsgaard, Jacob Sherson, J.~Ignacio Cirac, Jaromir Fiurasek, and
  Eugene~S. Polzik.
\newblock Experimental demonstration of quantum memory for light.
\newblock {\em Nature}, 432(7016):482--486, 2004.

\bibitem{Choi2008}
K.~S. Choi, H.~Deng, J.~Laurat, and H.~J. Kimble.
\newblock Mapping photonic entanglement into and out of a quantum memory.
\newblock {\em Nature}, 452(7183):67--71, 2008.

\bibitem{Liu2001}
Chien Liu, Zachary Dutton, Cyrus~H. Behroozi, and Lene~Vestergaard Hau.
\newblock Observation of coherent optical information storage in an atomic
  medium using halted light pulses.
\newblock {\em Nature}, 409(6819):490--493, 2001.

\bibitem{Lukin2001b}
D.~F. Phillips, A.~Fleischhauer, A.~Mair, R.~L. Walsworth, and M.~D. Lukin.
\newblock Storage of light in atomic vapor.
\newblock {\em Phys. Rev. Lett.}, 86:783--786, Jan 2001.

\bibitem{Murray2016}
C.~Murray and T.~Pohl.
\newblock Chapter seven - quantum and nonlinear optics in strongly interacting
  atomic ensembles.
\newblock In {\em Advances In Atomic, Molecular, and Optical Physics}, volume
  Volume 65, pages 321--372. Academic Press, 2016.

\bibitem{Pritchard2013a}
J.~D. {Pritchard}, K.~J. {Weatherill}, and C.~S. {Adams}.
\newblock {\em {Nonlinear Optics Using Cold Rydberg Atoms}}, pages 301--350.
\newblock World Scientific Publishing Co, 2013.

\bibitem{Dudin2012}
Y.~O. Dudin and A.~Kuzmich.
\newblock Strongly interacting rydberg excitations of a cold atomic gas.
\newblock {\em Science}, 336(6083):887--889, 2012.

\bibitem{Peyronel2012a}
Thibault Peyronel, Ofer Firstenberg, Qi-Yu Liang, Sebastian Hofferberth,
  Alexey~V. Gorshkov, Thomas Pohl, Mikhail~D. Lukin, and Vladan Vuletic.
\newblock Quantum nonlinear optics with single photons enabled by strongly
  interacting atoms.
\newblock {\em Nature}, 488(7409):57--60, 2012.

\bibitem{Gorniaczyk2014}
H.~Gorniaczyk, C.~Tresp, J.~Schmidt, H.~Fedder, and S.~Hofferberth.
\newblock Single-photon transistor mediated by interstate rydberg interactions.
\newblock {\em Phys. Rev. Lett.}, 113:053601, Jul 2014.

\bibitem{Tiarks2014}
Daniel Tiarks, Simon Baur, Katharina Schneider, Stephan D\"urr, and Gerhard
  Rempe.
\newblock Single-photon transistor using a f\"orster resonance.
\newblock {\em Phys. Rev. Lett.}, 113:053602, Jul 2014.

\bibitem{Kuzmich2000}
A.~Kuzmich, L.~Mandel, and N.~P. Bigelow.
\newblock Generation of spin squeezing via continuous quantum nondemolition
  measurement.
\newblock {\em Phys. Rev. Lett.}, 85:1594--1597, Aug 2000.

\bibitem{Wasilewski2010}
W.~Wasilewski, K.~Jensen, H.~Krauter, J.~J. Renema, M.~V. Balabas, and E.~S.
  Polzik.
\newblock Quantum noise limited and entanglement-assisted magnetometry.
\newblock {\em Phys. Rev. Lett.}, 104:133601, Mar 2010.

\bibitem{Leroux2010}
Ian~D. Leroux, Monika~H. Schleier-Smith, and Vladan Vuleti\'{c}.
\newblock Orientation-dependent entanglement lifetime in a squeezed atomic
  clock.
\newblock {\em Phys. Rev. Lett.}, 104:250801, Jun 2010.

\bibitem{Sewell2012}
R.~J. Sewell, M.~Koschorreck, M.~Napolitano, B.~Dubost, N.~Behbood, and M.~W.
  Mitchell.
\newblock Magnetic sensitivity beyond the projection noise limit by spin
  squeezing.
\newblock {\em Phys. Rev. Lett.}, 109:253605, Dec 2012.

\bibitem{Arecchi1965}
F.~Arecchi and R.~Bonifacio.
\newblock Theory of optical maser amplifiers.
\newblock {\em IEEE J. Quant. Electron.}, 1(4):169--178, 1965.

\bibitem{McCall1967}
S.~L. McCall and E.~L. Hahn.
\newblock Self-induced transparency by pulsed coherent light.
\newblock {\em Phys. Rev. Lett.}, 18:908--911, 1967.

\bibitem{Gorshkov2007}
Alexey~V. Gorshkov, Axel Andr\'e, Michael Fleischhauer, Anders~S. S\o{}rensen,
  and Mikhail~D. Lukin.
\newblock Universal approach to optimal photon storage in atomic media.
\newblock {\em Phys. Rev. Lett.}, 98:123601, Mar 2007.

\bibitem{Lester2015}
Brian~J. Lester, Niclas Luick, Adam~M. Kaufman, Collin~M. Reynolds, and
  Cindy~A. Regal.
\newblock Rapid production of uniformly filled arrays of neutral atoms.
\newblock {\em Phys. Rev. Lett.}, 115:073003, Aug 2015.

\bibitem{Barredo2016}
Daniel Barredo, Sylvain de~L{\'e}s{\'e}leuc, Vincent Lienhard, Thierry Lahaye,
  and Antoine Browaeys.
\newblock An atom-by-atom assembler of defect-free arbitrary two-dimensional
  atomic arrays.
\newblock {\em Science}, 354(6315):1021--1023, 2016.

\bibitem{Endres2016}
Manuel Endres, Hannes Bernien, Alexander Keesling, Harry Levine, Eric~R.
  Anschuetz, Alexandre Krajenbrink, Crystal Senko, Vladan Vuleti\'c, Markus
  Greiner, and Mikhail~D. Lukin.
\newblock Atom-by-atom assembly of defect-free one-dimensional cold atom
  arrays.
\newblock {\em Science}, 354(6315):1024--1027, 2016.

\bibitem{Haller2015}
Elmar Haller, James Hudson, Andrew Kelly, Dylan~A. Cotta, Bruno Peaudecerf,
  Graham~D. Bruce, and Stefan Kuhr.
\newblock Single-atom imaging of fermions in a quantum-gas microscope.
\newblock {\em Nat Phys}, 11(9):738--742, September 2015.

\bibitem{Greif2016}
Daniel Greif, Maxwell~F. Parsons, Anton Mazurenko, Christie~S. Chiu, Sebastian
  Blatt, Florian Huber, Geoffrey Ji, and Markus Greiner.
\newblock Site-resolved imaging of a fermionic mott insulator.
\newblock {\em Science}, 351(6276):953--957, 2016.

\bibitem{Zoubi2010}
H.~Zoubi and H.~Ritsch.
\newblock Metastability and directional emission characteristics of excitons in
  1d optical lattices.
\newblock {\em EPL (Europhysics Letters)}, 90(2):23001, 2010.

\bibitem{Bettles2015}
Robert~J. Bettles, Simon~A. Gardiner, and Charles~S. Adams.
\newblock Cooperative ordering in lattices of interacting two-level dipoles.
\newblock {\em Phys. Rev. A}, 92:063822, Dec 2015.

\bibitem{Bettles2016}
Robert~J. Bettles, Simon~A. Gardiner, and Charles~S. Adams.
\newblock Cooperative eigenmodes and scattering in one-dimensional atomic
  arrays.
\newblock {\em Phys. Rev. A}, 94:043844, Oct 2016.

\bibitem{Sutherland2016}
R.~T. Sutherland and F.~Robicheaux.
\newblock Collective dipole-dipole interactions in an atomic array.
\newblock {\em Phys. Rev. A}, 94:013847, Jul 2016.

\bibitem{Hebenstreit2017}
Martin Hebenstreit, Barbara Kraus, Laurin Ostermann, and Helmut Ritsch.
\newblock Subradiance via entanglement in atoms with several independent decay
  channels.
\newblock {\em Phys. Rev. Lett.}, 118:143602, Apr 2017.

\bibitem{Asenjo2017b}
A.~Asenjo-Garcia, M.~Moreno-Cardoner, A.~Albrecht, H.~J. Kimble, and D.~E.
  Chang.
\newblock Exponential improvement in photon storage fidelities using
  subradiance and ``selective radiance'' in atomic arrays.
\newblock {\em Phys. Rev. X}, 7:031024, Aug 2017.

\bibitem{Facchinetti2016}
G.~Facchinetti, S.~D. Jenkins, and J.~Ruostekoski.
\newblock Storing light with subradiant correlations in arrays of atoms.
\newblock {\em Phys. Rev. Lett.}, 117:243601, Dec 2016.

\bibitem{Perczel2017}
J.~Perczel, J.~Borregaard, D.~E. Chang, H.~Pichler, S.~F. Yelin, P.~Zoller, and
  M.~D. Lukin.
\newblock Topological quantum optics in two-dimensional atomic arrays.
\newblock {\em Phys. Rev. Lett.}, 119:023603, Jul 2017.

\bibitem{Bettles2017}
Robert~J. Bettles, Jiri Min\'a\ifmmode~\check{r}\else \v{r}\fi{}, Charles~S.
  Adams, Igor Lesanovsky, and Beatriz Olmos.
\newblock Topological properties of a dense atomic lattice gas.
\newblock {\em Phys. Rev. A}, 96:041603, Oct 2017.

\bibitem{Abajo2007}
F.~J. Garc\'{\i}a~de Abajo.
\newblock Colloquium.
\newblock {\em Rev. Mod. Phys.}, 79:1267--1290, Oct 2007.

\bibitem{Bettles2016b}
Robert~J. Bettles, Simon~A. Gardiner, and Charles~S. Adams.
\newblock Enhanced optical cross section via collective coupling of atomic
  dipoles in a 2d array.
\newblock {\em Phys. Rev. Lett.}, 116:103602, Mar 2016.

\bibitem{Shahmoon2017}
Ephraim Shahmoon, Dominik~S. Wild, Mikhail~D. Lukin, and Susanne~F. Yelin.
\newblock Cooperative resonances in light scattering from two-dimensional
  atomic arrays.
\newblock {\em Phys. Rev. Lett.}, 118:113601, Mar 2017.

\bibitem{Gross1982}
M.~Gross and S.~Haroche.
\newblock Superradiance: An essay on the theory of collective spontaneous
  emission.
\newblock {\em Physics Reports}, 93(5):301 -- 396, 1982.

\bibitem{Asenjo2017}
A.~Asenjo-Garcia, J.~D. Hood, D.~E. Chang, and H.~J. Kimble.
\newblock Atom-light interactions in quasi-one-dimensional nanostructures: A
  green's-function perspective.
\newblock {\em Phys. Rev. A}, 95:033818, Mar 2017.

\bibitem{Dung2002}
Ho~Trung Dung, Ludwig Kn\"oll, and Dirk-Gunnar Welsch.
\newblock Resonant dipole-dipole interaction in the presence of dispersing and
  absorbing surroundings.
\newblock {\em Phys. Rev. A}, 66:063810, Dec 2002.

\bibitem{Buhmann2007}
Stefan~Yoshi Buhmann and Dirk-Gunnar Welsch.
\newblock Dispersion forces in macroscopic quantum electrodynamics.
\newblock {\em Progr. in Quant. Electron.}, 31(2):51 -- 130, 2007.

\bibitem{GW96}
T.~Gruner and D.-G. Welsch.
\newblock Green-function approach to the radiation-field quantization for
  homogeneous and inhomogeneous {K}ramers-{K}ronig dielectrics.
\newblock 53:1818, 1996.

\bibitem{DKW02}
H.~T. Dung, L.~Kn{\"o}ll, and D.-G. Welsch.
\newblock Resonant dipole-dipole interaction in the presence of dispersing and
  absorbing surroundings.
\newblock 66:063810, 2002.

\bibitem{Scully2008}
S.~Das, G.~S. Agarwal, and M.~O. Scully.
\newblock Quantum interferences in cooperative {Dicke} emission from spatial
  variation of the laser phase.
\newblock {\em Phys. Rev. Lett.}, 101:15360, 2008.

\bibitem{Novotny2006}
Lukas Novotny and Bert Hecht.
\newblock {\em Principles of Nano-Optics}.
\newblock Cambridge University Press, 2006.

\bibitem{Meystre2007}
Pierre Meystre and Murray Sargent.
\newblock {\em Elements of Quantum Optics}.
\newblock Springer, 2007.

\bibitem{Scully2010}
A.~A. Svidzinsky, J.-T. Chang, and M.~O. Scully.
\newblock Cooperative spontaneous emission of {N} atoms: Many-body eigenstates,
  the effect of virtual lamb shift processes, and analogy with radiation of {N}
  classical oscillators.
\newblock {\em Phys. Rev. A}, 81:053821, 2010.

\bibitem{Ballestero2013}
C~Gonzalez-Ballestero, F~J Garcia-Vidal, and Esteban Moreno.
\newblock Non-markovian effects in waveguide-mediated entanglement.
\newblock {\em New Journal of Physics}, 15(7):073015, 2013.

\bibitem{Shi2015}
Tao Shi, Darrick~E. Chang, and J.~Ignacio Cirac.
\newblock Multiphoton-scattering theory and generalized master equations.
\newblock {\em Phys. Rev. A}, 92:053834, Nov 2015.

\bibitem{Guimond2016}
Pierre-Olivier Guimond, Alexandre Roulet, Huy~Nguyen Le, and Valerio Scarani.
\newblock Rabi oscillation in a quantum cavity: Markovian and non-markovian
  dynamics.
\newblock {\em Phys. Rev. A}, 93:023808, Feb 2016.

\bibitem{WSL04}
M.~Wubs, L.~G. Suttorp, and A.~Lagendijk.
\newblock Multiple-scattering approach to interatomic interactions and
  superradiance in inhomogeneous dielectrics.
\newblock {\em Phys. Rev. A}, 70:053823, 2004.

\bibitem{Suttorp2004}
L.~G. Suttorp and A.~J. van Wonderen.
\newblock Fano diagonalization of a polariton model for an inhomogeneous
  absorptive dielectric.
\newblock {\em EPL (Europhysics Letters)}, 67(5):766, 2004.

\bibitem{YVR09}
P.~Yao, C.~{Van Vlack}, A.~Reza, M.~Patterson, M.~M. Dignam, and S.~Hughes.
\newblock Ultrahigh {Purcell} factors and {Lamb} shifts in slow-light
  metamaterial waveguides.
\newblock {\em Phys. Rev. B}, 80:195106, 2009.

\bibitem{Manzonidouglas}
Marco~T Manzoni, Darrick~E Chang, and James~S Douglas.
\newblock {Simulating quantum light propagation through atomic ensembles using
  matrix product states}.
\newblock {\em Nature Communications}, 8(1):1743, 2017.

\bibitem{albrecht2017}
Andreas Albrecht, Tommaso Caneva, and Darrick~E Chang.
\newblock Changing optical band structure with single photons.
\newblock {\em New Journal of Physics}, 19(11):115002, 2017.

\bibitem{Bienias2018}
Przemyslaw Bienias, James Douglas, Asaf Paris-Mandoki, Paraj Titum, Ivan
  Mirgorodskiy, Christoph Tresp, Emil Zeuthen, Michael~J. Gullans, Marco
  Manzoni, Sebastian Hofferberth, Darrick Chang, and Alexey Gorshkov.
\newblock Photon propagation through dissipative rydberg media at large input
  rates.
\newblock {\em preprint at arXiv:1807.07586}, 2018.

\bibitem{Pletyukhov2012}
Mikhail Pletyukhov and Vladimir Gritsev.
\newblock Scattering of massless particles in one-dimensional chiral channel.
\newblock {\em New J. Phys.}, 14(9):095028, 2012.

\bibitem{Mahmoodian2018}
Sahand Mahmoodian, Mantas Cepulkovskis, Sumanta Das, Peter Lodahl, Klemens
  Hammerer, and Anders~S. S\o{}rensen.
\newblock Chiral waveguide qed: Strongly correlated photon transport with
  weakly coupled emitters.
\newblock {\em preprint at arXiv:1803.02428}, 2018.

\bibitem{Vetsch2010a}
E.~Vetsch, D.~Reitz, G.~Sagu\'e, R.~Schmidt, S.~T. Dawkins, and
  A.~Rauschenbeutel.
\newblock Optical interface created by laser-cooled atoms trapped in the
  evanescent field surrounding an optical nanofiber.
\newblock {\em Phys. Rev. Lett.}, 104:203603, 2010.

\bibitem{KGB05}
F.~{Le Kien}, S.~D. Gupta, V.~I. Balykin, and K.~Hakuta.
\newblock Spontaneous emission of a cesium atom near a nanofiber: efficient
  coupling of light to guided modes.
\newblock 72:032509, 2005.

\bibitem{Hood2016}
Jonathan~D. Hood, Akihisa Goban, Ana Asenjo-Garcia, Mingwu Lu, Su-Peng Yu,
  Darrick~E. Chang, and H.~J. Kimble.
\newblock Atom-atom interactions around the band edge of a photonic crystal
  waveguide.
\newblock {\em Proc. Natl. Acad. Sci. U.S.A.}, 113(38):10507--10512, 2016.

\bibitem{Lalumiere2013}
Kevin Lalumi\`ere, Barry~C. Sanders, A.~F. van Loo, A.~Fedorov, A.~Wallraff,
  and A.~Blais.
\newblock Input-output theory for waveguide qed with an ensemble of
  inhomogeneous atoms.
\newblock {\em Phys. Rev. A}, 88:043806, 2013.

\bibitem{LFL13}
A.~F. {van Loo}, A.~Fedorov, K.~Lalumiere, B.~C. Sanders, A.~Blais, and
  A.~Wallraff.
\newblock Photon-mediated interactions between distant artificial atoms.
\newblock {\em Science}, 342:1494--1496, 2013.

\bibitem{maxwell}
D.~Maxwell, D.~J. Szwer, D.~Paredes-Barato, H.~Busche, J.~D. Pritchard,
  A.~Gauguet, K.~J. Weatherill, M.~P.~A. Jones, and C.~S. Adams.
\newblock Storage and control of optical photons using rydberg polaritons.
\newblock {\em Phys. Rev. Lett.}, 110:103001, Mar 2013.

\bibitem{sipahigil}
A.~Sipahigil, K.~D. Jahnke, L.~J. Rogers, T.~Teraji, J.~Isoya, A.~S. Zibrov,
  F.~Jelezko, and M.~D. Lukin.
\newblock Indistinguishable photons from separated silicon-vacancy centers in
  diamond.
\newblock {\em Phys. Rev. Lett.}, 113:113602, Sep 2014.

\bibitem{Labuhn2016}
Henning Labuhn, Daniel Barredo, Sylvain Ravets, Sylvain de~L\'es\'eleuc,
  Tommaso Macr\`i, Thierry Lahaye, and Antoine Browaeys.
\newblock Tunable two-dimensional arrays of single rydberg atoms for realizing
  quantum ising models.
\newblock {\em Nature}, 534(7609):667--670, June 2016.

\bibitem{Chen2002}
Carl~G. Chen, Paul~T. Konkola, Juan Ferrera, Ralf~K. Heilmann, and Mark~L.
  Schattenburg.
\newblock Analyses of vector gaussian beam propagation and the validity of
  paraxial and spherical approximations.
\newblock {\em J. Opt. Soc. Am. A}, 19(2):404--412, Feb 2002.

\end{thebibliography}
\vfill

\end{document}